\documentclass[preprint,prd,aps,floats,amssymb,showpacs,draft,%
floatfix,superscriptaddress,nofootinbib]{revtex4} 
\usepackage{psfig}

\begin{document}

\title{ 
Strong-disorder paramagnetic-ferromagnetic fixed point \\ 
in the square-lattice $\pm J$ Ising model.
}

\author{Francesco Parisen Toldin} 
\affiliation{ 
Max-Planck-Institut f\"ur Metallforschung,
Heisenbergstrasse 3, D-70569 Stuttgart, Germany\\
and Institut f\"ur Theoretische und Angewandte Physik,
Universit\"at Stuttgart,
Pfaffenwaldring 57, D-70569 Stuttgart, Germany.
} 
\author{Andrea Pelissetto} 
\affiliation{Dipartimento di Fisica
  dell'Universit\`a di Roma ``La Sapienza" and INFN, \\
  Piazzale Aldo Moro 2, I-00185 Roma, Italy.}
\author{Ettore Vicari} 
\affiliation{ 
Dipartimento di Fisica dell'Universit\`a di Pisa and INFN,\\
Largo Pontecorvo 3, I-56127 Pisa, Italy.  } 

\date{\today}

\begin{abstract}
  We consider the random-bond $\pm J$ Ising model on a square lattice as a
  function of the temperature $T$ and of the disorder parameter $p$ ($p=1$
  corresponds to the pure Ising model).  We investigate the critical behavior
  along the paramagnetic-ferromagnetic transition line at low temperatures,
  below the temperature of the multicritical Nishimori point 
  at $T^*= 0.9527(1)$, $p^*=0.89083(3)$. We
  present finite-size scaling analyses of Monte Carlo results at two
  temperature values, $T \approx 0.645$ and $T=0.5$.  The results show that
  the paramagnetic-ferromagnetic transition line is reentrant for $T<T^*$,
  that the transitions are continuous and controlled by a strong-disorder
  fixed point with critical exponents $\nu = 1.50(4)$, $\eta = 0.128(8)$,
  and $\beta = 0.095(5)$.
  This fixed point is definitely different from the Ising fixed point 
  controlling the paramagnetic-ferromagnetic transitions for $T>T^*$.
  Our results for the critical exponents are consistent with the 
  hyperscaling relation $2\beta/\nu - \eta = d - 2 = 0$.
\end{abstract}

\pacs{75.10.Nr, 64.60.Fr, 75.40.Cx, 75.40.Mg}


\maketitle


\section{Introduction}
\label{intro}

The $\pm J$ Ising model represents an interesting theoretical laboratory, in
which one can study the effects of quenched disorder and frustration on the
critical behavior of spin systems. While originally introduced to describe
magnetic systems with disordered couplings \cite{EA-75}, it has been shown
recently to be also relevant for quantum
computations~\cite{DKLP-02,Kitaev-03}.  It is defined by the lattice
Hamiltonian~\cite{EA-75}
\begin{equation}
{\cal H} = - \sum_{\langle xy \rangle} J_{xy} \sigma_x \sigma_y,
\label{lH}
\end{equation}
where $\sigma_x=\pm 1$, the sum is over all pairs of lattice nearest-neighbor
sites, and the exchange interactions $J_{xy}$ are uncorrelated quenched random
variables, taking values $\pm J$ with probability distribution
\begin{equation}
P(J_{xy}) = p \delta(J_{xy} - J) + (1-p) \delta(J_{xy} + J). 
\label{probdis}
\end{equation}
In the following we set $J=1$ without loss of generality.  For $p=1$ we
recover the standard Ising model, while for $p=1/2$ we obtain the bimodal
Ising spin-glass model.

\begin{figure*}[tb]
\centerline{\psfig{width=9truecm,angle=0,file=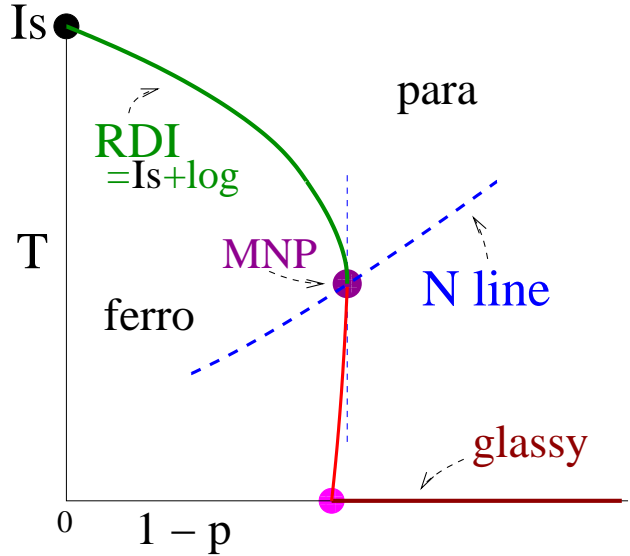}}
\caption{(Color online)
Phase diagram of the square-lattice $\pm J$ Ising model for 
$1-p\le 1/2$. }
\label{phdia}
\end{figure*}

The $T$-$p$ phase diagram of the two-dimensional (2D) square-lattice $\pm J$
Ising model has been extensively investigated 
\cite{Nishimori-01,KR-03,
  Hartmann-08,KLC-07,JLMM-06,AMMP-03,HY-01,
  HPPV-08-2,BGP-98,Nishimori-81,Ohzeki-08,
  HPPV-08,Nishimori-07,Nishimori-05,PHP-06,Queiroz-06,QS-03,MNN-03,MC-02,
  NFO-02,NN-02,HPP-01,
  Nobre-01,GRL-01,AQS-99,OI-98,MB-98,CF-97,Simkin-97,SA-96,ON-93,
  Kitatani-92,LH-89,ON-87,Nishimori-86,GHDB-85,GHDMB-86,McMillan-84,
  AH-04,WHP-03,KR-97}.  The resulting phase diagram, which is sketched in
Fig.~\ref{phdia}, presents two phases at finite temperature: a paramagnetic
and a ferromagnetic phase.  They are separated by a transition line, which
starts at the pure Ising transition point at $p=1$ and $T_{\rm Is}\approx
2.269$ and ends at the $T=0$ transition at $p_0\approx 0.897$. 
 The point where this
transition line meets the so-called Nishimori (N) line \cite{Nishimori-81}, at
$T^*=0.9527(1)$ and $p^*=0.89083(3)$ (we derive these 
estimates in the present paper), is a multicritical point (MNP) \cite{LH-89}.

The MNP divides the paramagnetic-ferromagnetic (PF) transition line in two
parts.  The PF transition line from the Ising point at $p=1$ to the MNP is
controlled by the Ising fixed point: disorder gives only rise to logarithmic
corrections to the standard Ising critical behavior \cite{HPPV-08-2}.  On the
other hand, the presence of the MNP on the transition line suggests that the
PF transitions for $T<T^*$ belong to a different strong-disorder universality
class. This is confirmed by the renormalization-group (RG) calculations of
Refs.~\onlinecite{McMillan-84} and \onlinecite{MB-98}, using domain-wall and
Migdal-Kadanoff RG transformations respectively, which found that the RG flow
along the critical line for $T<T^*$ was attracted by a different fixed point.

In this paper we investigate the critical behavior along the low-temperature
transition line from the MNP to the $T=0$ axis.  We perform Monte Carlo (MC)
simulations at two temperature values below the MNP, i.e., at $\beta \equiv
1/T = 2$ and $\beta=1.55$.  As we shall see, our finite-size scaling (FSS)
analyses show that the PF transition line for $T<T^*$ is reentrant and that
the transitions are continuous. Moreover, the estimates of the critical
exponents and of several RG invariant quantities for these two values of $T$
are consistent, supporting the hypothesis that the PF transition line below
the MNP belongs to a unique universality class.  The values of the critical
exponents, $\nu = 1.50(4)$, $\eta = 0.128(8)$, and $\beta = 0.095(5)$ are
clearly different from the Ising values $\nu = 1$, $\eta = 1/4$, $\beta =
1/8$.  Therefore, these results show the existence of a strong-disorder fixed
point associated with a PF transition.  Note that this strong-disorder fixed
point does not violate hyperscaling.  Indeed, our results are consistent with
the hyperscaling relation $2+2\beta/\nu - \eta = d = 2$ (our estimates of the
critical exponents $\eta$ and $\beta$ give $2 + 2\beta/\nu - \eta = 2.00(1)$).
The transitions for $T<T^*$ are no longer in the basin of attraction of the
Ising fixed point, which is the relevant one for small disorder and determines
the critical behavior along the transition line for $T>T^*$.

The paper is organized as follows.  In Sec.~\ref{phasediagram} we review the
main features of the $T$-$p$ phase diagram of the square-lattice $\pm J$ Ising
model.  The MC results and their FSS analyses are presented in
Sec.~\ref{MCsec}. In Sec.~\ref{conclusions} we draw our conclusions.  Some
technical details on the simulations are presented in App.~\ref{MC-details},
while the quantities we compute are defined in App.~\ref{notations}. In
App.~\ref{MNP} we present a reanalysis of the critical behavior at the MNP,
using the additional data we have collected in this work. Moreover, we also
present analyses which take into account the analytic corrections, which had
been neglected in our previous work \cite{HPPV-08}.  This allows us to obtain
improved estimates of the critical parameters at the MNP.

\section{The phase diagram of the square-lattice $\pm J$ Ising 
model} 
\label{phasediagram}

The phase diagram of the square-lattice $\pm J$ Ising model is sketched in
Fig.~\ref{phdia}.  It is symmetric for $p\rightarrow 1-p$ and thus we only
report it for $1-p\le 1/2$.  For sufficiently small values of the probability
of antiferromagnetic bonds $p_a\equiv 1-p$, the model presents a paramagnetic
phase and a ferromagnetic phase, separated by a transition line.  The PF
transition line starts at the Ising point $X_{\rm Is}=(T=T_{\rm Is},p=1)$,
where $T_{\rm Is}=2/\ln(1+\sqrt{2})=2.26919...$ is the critical temperature of
the 2D Ising model, and extends up to a $T=0$ transition
at~\cite{AH-04,WHP-03} $X_0=(T=0,p=p_0\approx 0.897)$.

The slope of the transition line at $p=1$ is known exactly \cite{Domany-79},
so that for small $1-p$ we have
\begin{equation}
T_c(p) = T_{\rm Is} \left[1 - {2 \sqrt{2}\over \ln(1+\sqrt{2})} (1 - p) + 
    \ldots \right].
\label{TCder}
\end{equation}
In the $T$-$p$ phase diagram an important role is played by the Nishimori (N)
line~\cite{Nishimori-81,Nishimori-01} defined by the equation
($p\ge 1/2$)
\begin{equation}
T = T_N(p),
\qquad 
T_N(p) = {2\over \ln p - \ln(1-p)}. 
\label{tn}
\end{equation}
Along the N-line several rigorous results can be proved
\cite{Nishimori-81,Nishimori-02,Nishimori-01}.  The energy density is given by
\begin{equation}
E_N(p) \equiv {1\over V} [ \langle {\cal H} \rangle_{T_N(p)} ] = 2-4p,
\label{energy}
\end{equation} 
and the spin-spin
and the overlap correlation functions are equal
\begin{equation}
[\langle \sigma_0 \sigma_x \rangle] = [\langle \sigma_0 \sigma_x
\rangle^2].
\label{eqfunc}
\end{equation} 
Here the angular and square brackets refer respectively to the thermal average
and to the quenched average over the bond couplings $\{J_{xy}\}$.  As argued
in Refs.~\cite{GHDB-85,LH-89} and verified numerically
\cite{HPPV-08,PHP-06,Queiroz-06,QS-03,MC-02,SA-96}, the critical point $X_{\rm
  MNP}=(T^*\approx 0.953,p^*\approx 0.891)$ along the N line is a
multicritical point (MNP).

Along the transition line from the Ising point $X_{\rm Is}$ to the MNP, the
critical behavior is analogous to that observed in 2D randomly dilute Ising
(RDI) models~ \cite{HPPV-08-2}. It is controlled by the pure Ising fixed point
and disorder is marginally irrelevant, giving rise to a universal pattern of
logarithmic corrections, see, e.g.,
Refs.~\cite{HPPV-08-2,Shalaev-84,Shankar-87,LC-87} and references therein.

The location of the MNP and the corresponding critical exponents can be
obtained by FSS analyses of MC data along the N line. The new analysis
reported in App.~\ref{MNP} gives
\begin{equation}
T^*=0.9527(1), \qquad p^*=0.89083(3).
\label{mnpest}
\end{equation}
In the absence of external fields, the MNP is characterized by two relevant RG
operators with RG dimensions $y_1=0.66(1)$ and $y_2=0.250(2)$. Moreover, the
magnetic exponent $\eta$ is given by $\eta = 0.177(2)$.  Other estimates of
$T^*$, $p^*$, and of the critical exponents can be found in
Refs.~\cite{MC-02,PHP-06,Nishimori-07,HPPV-08,Ohzeki-08}.

As a consequence of the inequality~\cite{Nishimori-81}
\begin{equation}
|[\langle\sigma_x\sigma_y\rangle_T]_p|\le
[|\langle\sigma_x\sigma_y\rangle_{T_N(p)}|]_p
\label{ineq}
\end{equation}
(the subscripts indicate the values of $T$ and $p$ at which the thermal and
disorder average are performed), ferromagnetism can only exist in the region
$p\ge p^*$. Thus, the PF boundary lies in the region $p\ge p^*$ and, at the
MNP, the transition line is tangent to the line $p = p^*$, hence parallel to
the $T$ axis. As a further consequence, at $T=0$ the ferromagnetic phase ends
at $p=p_0$ with $p_0\ge p^*$.  In
Refs.~\cite{ON-93,Kitatani-92,Nishimori-86,Nishimori-01} it was argued that
the PF transition line from the MNP to $X_0=(0,p_0)$ is only related to the
frustration distribution; hence, it should not depend on temperature and
should coincide with the line $p=p^*$, so that $p_0=p^*$.  Numerical estimates
of $p_0$ have shown that this argument is not exact.  Indeed, numerical
analyses \cite{PHP-06,AH-04,WHP-03,MC-02,BGP-98,KR-97} give $p_0\approx
0.897$;\footnote{The most precise estimates are apparently \cite{AH-04}
  $p_0=0.897(1)$ and \cite{WHP-03} $p_0=0.8969(1)$.}  this suggests that the
transition line below the MNP is reentrant, i.e.  $p_c>p^*$ for any $T<T^*$.
The difference is however quite small, $p_0-p^*\approx 0.006$.

Our FSS analyses confirm that the PF transition line is reentrant for $T<T^*$.
Indeed, we find $p_c=0.8915(2)$ at $T=1/1.55\approx 0.645$ and $p_c=0.8925(1)$
at $T=0.5$.  The PF transitions are of second order and show the same critical
behavior with critical exponents $\nu = 1.50(4)$, $\eta = 0.128(8)$, and
$\beta = 0.095(5)$, which are consistent with hyperscaling.  These results
confirm the existence of a strong-disorder fixed point, different from the
Ising fixed point which controls the PF transitions above the MNP, i.e. for
$T^*<T<T_{\rm Is}$.

At variance with the three-dimensional case, there is no evidence of a
finite-temperature glassy phase.  Glassy behavior is only expected for $T=0$
and $p<p_0$.  The critical behavior for $T\to 0$ has been much investigated
for $p=1/2$ \cite{Hartmann-08,KLC-07,JLMM-06,AMMP-03,HY-01}. In particular,
simulations found that the correlation length increases as $T^{-\nu}$ with
$\nu\approx 3.5$.  A natural hypothesis is that a $T=0$ glassy transition
occurs for any $p<p_0$, with critical behavior in the same universality class
as that of the bimodal model with $p=1/2$.

The point $X_0=(0,p_0)$, where the low-temperature transition line ends is a
multicritical point: it is connected to three phases and it is the
intersection of two different transition lines, the PF line at $T>0$ and the
glassy line at $T=0$.  At $T=0$ the critical point $X_0$ separates a
ferromagnetic phase from a $T=0$ glassy phase, while for $T > 0$ the
transition line separates a ferromagnetic from a paramagnetic phase.
Therefore, on general grounds, the critical behavior when varying $p$ at $T=0$
differs from that along the PF transition line at $T>0$, unless the magnetic
and glassy critical modes are effectively decoupled at the $T=0$ multicritical
point. The latter scenario is apparently supported by the fact that the
estimates of magnetic critical exponents at $T=0$, see e.g.
Refs.~\onlinecite{McMillan-84,WHP-03,AH-04,PHP-06}, are quite close and
substantially consistent with those found along the transition line at finite
temperature $0<T<T^*$.

\section{Monte Carlo results}
\label{MCsec}

We investigate the critical behavior along the PF line
that starts at the MNP $T^*\approx 0.95$ and ends at $T = 0$.  Since the
transition line below the MNP is expected to be almost parallel to the $T$
axis, we study the FSS behavior of several quantities at fixed $T$ as a
function of $p$. We consider two values of $T$, $\beta\equiv 1/T=2$ and
$\beta=1.55$, which are quite far from the two endpoints of the line.
For each of these two values we perform MC simulations on square lattices of
linear size $L$ with periodic boundary conditions, for several values of $L$:
$L=8,12,16,24,32,48,64$.  In our MC simulations we employ the Metropolis
algorithm, the random-exchange method (often called parallel-tempering or
multiple Markov-chain method) \cite{raex,par-temp}, and multispin coding.
Some details are reported in App.~\ref{MC-details}.

\subsection{The critical point $p_c$ and exponent $\nu$}

\begin{figure*}[tb]
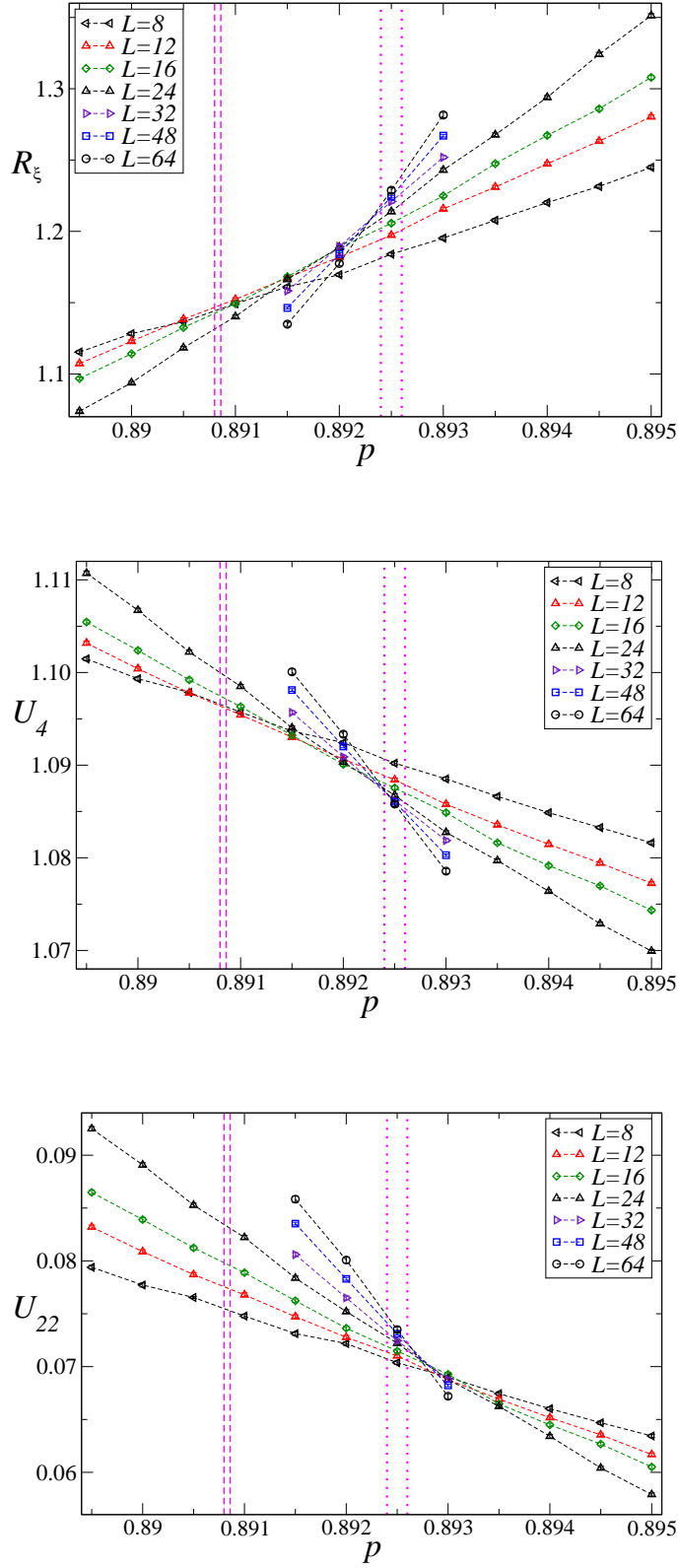

\centerline{\psfig{width=9truecm,angle=0,file=rxi.eps}}
\vspace{12mm}
\centerline{\psfig{width=9truecm,angle=0,file=U4.eps}}
\vspace{12mm}
\centerline{\psfig{width=9truecm,angle=0,file=U22.eps}}
\caption{(Color online) MC estimates of $R_\xi\equiv \xi/L$, $U_4$, $U_{22}$
  at $\beta=2$ vs $p$.  The lines connecting the data at given $L$ are drawn
  to guide the eye.  The dashed vertical lines corresponds to the MNP location
  $p^*=0.89083(3)$.  The dotted vertical lines indicate our final estimate of
  $p_c$, $p_c=0.8925(1)$.}
\label{rgifig}
\end{figure*}

We first focus on the data at $\beta = 2$, for which we
have most of the statistics.  In order to estimate $p_c$ and $\nu$, we perform
a FSS analysis of the renormalized couplings $R_\xi$, $U_4$, $U_{22}$, which
are defined in App.~\ref{notations} and are generically denoted by $R$ in the
following.  MC estimates are shown in Fig.~\ref{rgifig}.  We
clearly observe a crossing point for $0.8920 < p < 0.8930$, indicating $p_c
\approx 0.892$-0.893, which is larger than the value at the MNP, i.e. $p_c =
p^*=0.89083(3)$. This already suggests that the transition line
is reentrant. 

To obtain more precise estimates we perform a careful FSS analysis, following
Ref.~\cite{HPV-08-SG}. 
In the FSS limit any RG invariant quantity obeys the scaling law
\begin{equation}
R = f_R( u_1 L^{y_1}),
\label{FSS-R}
\end{equation}
where $f_R(0) = R^*$, $y_1 \equiv 1/\nu$, and we have neglected scaling
corrections.  Here $u_1$ is the nonlinear scaling field associated with the
leading relevant operator, which has RG dimension $y_1$. The scaling field is
an analytic function of the system parameters which vanishes along the
critical line. Thus, for $p\to p_c(\beta)$ at fixed $\beta$ we can write
\begin{equation}
    u_1 = A_0(\beta) (p - p_c) + A_1(\beta) (p - p_c)^2 + \ldots
\label{exp-u1}
\end{equation}
where the coefficients $A_i(\beta)$ are analytic functions of $\beta$.  The
terms of order $(p - p_c)^2,(p - p_c)^3$, etc., give rise to
corrections of order $L^{-n y_1}$ as $L\to\infty$. They are named {\rm 
analytic} corrections, because they arise from the
analytic dependence of the scaling fields on the model parameters.  See
Ref.~\cite{HPV-08-SG} for a thorough discussion of their origin. In pure
ferromagnetic systems, in which $\nu \lesssim 1$ and 
$y_1 \gtrsim 1$, they are usually
negligible, and the nonanalytic corrections, which behave as $L^{-\omega}$,
$\omega \lesssim 1$, play a much more important role. This is not the case
here, since, as we shall see, at the transition line $y_1\approx \omega < 1$.

Since our data are sufficiently close to the critical point, $p - p_c$ is
small and thus we can take $u_1 \sim (p - p_c)$. Moreover, also 
the product $(p -
p_c) L^{y_1}$ is small, so that we can expand $f_R(x)$ in powers of $x$. Thus,
we fit the numerical data to
\begin{equation}
   R = R^* + \sum_{n=1}^{n_{\rm max}} a_n (p - p_c)^n L^{n y_1}, 
\label{fitR-1}
\end{equation}
keeping $R^*$, the coefficients $\{a_n\}$, $p_c$, and $y_1$ as free
parameters.  Here we neglect scaling corrections.  To monitor their role, we
repeat the fits several times, each time only including data satisfying $L\ge
L_{\rm min}$. For a given $L_{\rm min}$, $\chi^2/{\rm DOF}$ (DOF is the number
of degrees of freedom of the fit) changes significantly as we increase $n_{\rm
  max}$ from 1 to 2, and only marginally as we change this parameter from 2 to
3.  This indicates that the range of values of $p$ we are considering is too
large to allow for a linear approximation of the scaling function $f_R(x)$.
Instead, a quadratic approximation seems to be accurate enough.  Thus, the
results we present below correspond to $n_{\rm max} = 2$.

\begin{table}
  \begin{tabular}{lcclllll}
\hline
\hline
\multicolumn{1}{c}{$L_{\rm min}$}&
\multicolumn{1}{c}{$\chi^2/$DOF}&
\multicolumn{1}{c}{$\omega$}&
\multicolumn{1}{c}{$R_\xi^*$}&
\multicolumn{1}{c}{$U_4^*$}&
\multicolumn{1}{c}{$U_{22}^*$}&
\multicolumn{1}{c}{$p_c$} &
\multicolumn{1}{c}{$y_1$} \\
\hline
8  & 9846/166 &      & 1.1865(3) &  1.09028(5) & 0.07287(4)  &
       0.892163(5) &   0.674(3) \\
12 & 3689/130 &      & 1.1996(4) &  1.08867(6) & 0.07293(5)  &
       0.892294(6) &   0.678(4) \\
16 & 1522/94  &      & 1.2068(4) &  1.08792(8) & 0.07327(7)  &
       0.892348(8) &   0.673(6) \\
24 & 441/61   &      & 1.2129(8) &  1.08734(12)& 0.07355(11) &
       0.892389(11)&   0.676(8) \\
32 &  96/25   &      & 1.2172(16)&  1.08695(22)& 0.07350(20) &
       0.892431(19)&   0.661(21)\\ \hline
8  &  212/159 & 0.58(4) &1.265(5)& 1.0810(6) & 0.0724(2) &
     0.89265(2) & 0.677(17) \\
12 &  96/123  & 0.64(8) &1.249(6)& 1.0836(7) & 0.0741(3) &
     0.89254(3) & 0.667(22) \\
16 &  77/87   & 0.63(13)&1.246(9)& 1.0842(9) & 0.0746(5) &
     0.89251(3) & 0.660(30) \\
24 &  48/54   & 0.50(43)&1.254(38)& 1.0840(37) & 0.0752(25) &
     0.89251(11) & 0.46(12) \\
\hline
\hline
  \end{tabular}
  \caption{Estimates obtained from the analysis of the data at $\beta = 2$.
   Above we report the results of the combined fits of $R_\xi$, $U_4$,
   and $U_{22}$ to Eq.~(\ref{fitR-1}) with $n_{\rm max} = 2$. 
   Below we report the results of the fits to Eq.~(\ref{fitR-2}) 
   with $n_{\rm max} = 2$ and $k_{\rm max} = 1$.
   }
  \label{res-y1-beta2}
\end{table}

In Table \ref{res-y1-beta2} we give the estimates of $R^*$, $p_c$, and $y_1$
from combined fits of $R_\xi$, $U_4$, and $U_{22}$.  All quantities,
except $y_1$, show a significant---much larger than the statistical
errors---variation with $L_{\rm min}$. Moreover, the $\chi^2$ is very
large. Clearly, scaling corrections are not negligible.  In order to take them
into account, we fit the MC data to
\begin{equation}
   R = R^* + \sum_{n=1}^{n_{\rm max}} a_n (p-p_c)^n L^{ny_1} + 
       L^{-\omega} \sum_{k=0}^{k_{\rm max}} b_k (p-p_c)^k L^{ky_1},
\label{fitR-2}
\end{equation}
taking $\omega$ as a free parameter.
Results for $k_{\rm max} = 1$ and $n_{\rm max} = 2$ are also 
reported in Table  \ref{res-y1-beta2}. The $\chi^2$ is now significantly
smaller and $\chi^2/{\rm DOF}\approx 1$, indicating that the fitting form
(\ref{fitR-2}) describes the data at the level of  their
statistical accuracy. The results are stable and the estimates for 
$L_{\rm min} \ge 12$ are consistent within errors. These fits also provide an 
estimate of the correction-to-scaling exponent $\omega$. We find
\begin{eqnarray}
  \omega = 0.6(1).
\end{eqnarray}
The estimates of $\omega$ and $y_1$ indicate that $\omega \approx y_1$, 
so that analytic and nonanalytic corrections behave analogously. Therefore,
we should also consider the analytic corrections.
For this purpose, we also performed fits to 
\begin{equation}
   R = R^* + \sum_{n=1}^{n_{\rm max}} a_n [1 + c (p-p_c)]^n 
   (p - p_c)^n L^{n y_1}, 
\label{fitR-3}
\end{equation}
which corresponds to including the quadratic term in the expansion of the 
nonlinear scaling field $u_1$. The parameter $c$ is a new fitting parameter
which is independent of the quantity one is analyzing. 
Fits to Eq.~(\ref{fitR-3}) are substantially equivalent to those to 
Eq.~(\ref{fitR-1}). For instance, the $\chi^2$ of the combined fit 
for $L_{\rm min} = 8$ is 9846, which is identical to that 
reported in Table~\ref{res-y1-beta2} for the same 
value of $L_{\rm min}$. The coefficient $c$ is small and we 
estimate $|c| \lesssim 0.3$. Since our data satisfy $|p - p_c|\le 0.0030$,
the analytic term gives a tiny correction and does not influence the 
fit results. 

Comparing the results of the different fits we arrive at the final estimates
\begin{eqnarray}
 y_1 &=& 0.67(2), \qquad \nu = 1/y_1 = 1.50(4), \label{est-y1}\\
 p_c &=& 0.8925(1),                    \label{est-pc-beta2} \\
 R^*_\xi &=& 1.25(3),  \label{Rxi-beta2} \\
 U^*_4 &=& 1.084(3),   \label{U4-beta2} \\
 U^*_{22} &=& 0.074(1) . \label{U22-beta2} 
\end{eqnarray}
The central value corresponds to the result of the fit to 
Eq.~(\ref{fitR-2}) with $L_{\rm min} = 12$; the errors are such to include
the results of the fits to Eq.~(\ref{fitR-1}) and 
$L_{\rm min} = 32$, and should take into account
the systematic error due to further scaling corrections
which have been neglected in our analyses.

\begin{table}
  \begin{tabular}{lcclllll}
\hline
\hline
\multicolumn{1}{c}{$L_{\rm min}$}&
\multicolumn{1}{c}{$\omega$}&
\multicolumn{1}{c}{$\chi^2/$DOF}&
\multicolumn{1}{c}{$R_\xi^*$}&
\multicolumn{1}{c}{$U_4^*$}&
\multicolumn{1}{c}{$U_{22}^*$}&
\multicolumn{1}{c}{$p_c$} &
\multicolumn{1}{c}{$y_1$} \\
\hline
8  &   & 2099/88 & 1.1362(8) &   1.0984(1) & 0.0756(1) &
    0.89107(2) &  0.580(10) \\
12 &   & 1045/73 & 1.1457(8) &   1.0969(1) & 0.0753(1) &
    0.89122(2) &  0.626(12) \\
16 &   &  556/58 & 1.1508(10)&   1.0962(2) & 0.0753(1) &
    0.89128(2) &  0.636(13) \\
24 &   &  236/43 & 1.1543(15)&   1.0959(3) & 0.0755(2) &
    0.89131(3) &  0.623(17) \\
\hline
8 & 0.4&   77/82 & 1.213(3) &    1.0878(5) & 0.0739(4) &
    0.89160(3) &  0.663(52) \\
8 & 0.6&   88/82 & 1.190(2) &    1.0908(4) & 0.0744(3) &
    0.89153(2) &  0.662(37) \\
8 & 0.8&  111/82 & 1.178(2) &    1.0904(3) & 0.0746(2) &
    0.89148(2) &  0.659(29) \\
\hline
\hline
  \end{tabular}
  \caption{Estimates obtained from the analysis of the data at $\beta = 1.55$.
   Above we report the results of the combined fits of $R_\xi$, $U_4$,
   and $U_{22}$ to Eq.~(\ref{fitR-1}) with $n_{\rm max} = 2$. 
   Below we report the results of fits to Eq.~(\ref{fitR-2}) 
   with $n_{\rm max} = 2$, $k_{\rm max} = 1$, and 
   $\omega$ fixed to 0.4, 0.6, 0.8.
    }
  \label{res-y1-beta1p55}
\end{table}

We repeat the same type of analysis at $\beta = 1.55$. We report in
Table~\ref{res-y1-beta1p55} the results of the fits to Eqs.~(\ref{fitR-1}) and
(\ref{fitR-2}). In the latter case the data do not allow us to perform fits
in which $\omega$ is a free parameter.  Thus, we
only report results of fits in which $\omega$ is fixed to 0.4, 0.6, and 0.8,
consistently with the estimate $\omega \approx 0.6$ presented above. Fits
without scaling corrections are characterized by large values of $\chi^2$/DOF
and by a systematic trend of the results.  Fits with scaling corrections are
significantly better.  The estimates of $y_1$ and $U_{22}^*$ are in perfect
agreement with those obtained at $\beta = 2$. Those of $U_4^*$ and $R_\xi^*$
are substantially consistent: the difference between the estimates
(\ref{Rxi-beta2}), (\ref{U4-beta2}) and the results of the fit with $\omega =
0.6$---this is the fit which, in principle, should be more reliable---is of
the order of two error bars and can thus be explained by the presence of
residual scaling corrections which are not taken into account in our 
error estimate. Therefore, our analyses of the renormalized
couplings are consistent with a critical transition line whose nature is $T$
independent: for $T < T^*$, the PF transition belongs to a unique universality
class.

The estimate (\ref{est-y1}) is different from the Ising value $\nu=1$.
Therefore, the PF fixed point associated with the transitions along the line
$T<T^*$ is a new one, clearly distinct from the Ising one, which controls the
critical behavior for weak disorder. Analogously, our estimates of the
critical value of the renormalized couplings differ from the Ising values
\cite{SS-00,HPPV-08-2} $R_\xi^* = 0.9050488292(4)$, $U_4^* = 1.167923(5)$,
$U_{22}^* = 0$, and from those at the MNP (see App.~\ref{MNP}), 
which are $R_\xi^* =
0.997(1)$, $U_4^* = 1.1264(4)$, and $U_{22}^* = 0.0817(3)$.

Our analyses also give an estimate of $p_c$ for $\beta = 1.55$:
\begin{equation}
p_c = 0.8915(2) .
   \label{est-pc-beta1.55}
\end{equation}
Therefore, for  both values of $\beta$ we find
$p_c > p^*=0.89083(3)$. 
Thus, the PF transition line is reentrant,
contradicting the conjecture of 
Refs.~\cite{ON-93,Kitatani-92,Nishimori-86,Nishimori-01}.

\subsection{The exponent $\eta$}

We determine the critical exponent $\eta$ from the critical 
behavior of the susceptibility $\chi$. As discussed in Ref.~\cite{HPV-08-SG}
in the context of the three-dimensional paramagnetic-glassy transition, 
close to the critical point the susceptibility $\chi$ behaves as 
\begin{equation}
\chi = \bar{u}_h^2 L^{2-\eta} f_\chi[(p - p_c) L^{y_1}],
\label{FSS-chi}
\end{equation}
where $\bar{u}_h$ is a function of $p$ related to the magnetic 
nonlinear scaling field. Note that we have approximated 
$u_1$ with $p - p_c$, because, as already discussed, the 
analytic dependence of the scaling field $u_1$ is negligible for our data.

Since we are very close to the critical point, we can expand all 
quantities in powers of $(p - p_c)$. For this reason we 
perform fits to 
\begin{equation}
\ln \chi = (2-\eta) \ln L + \sum_{n=0}^{n_{\rm max}} 
    a_n (p - p_c)^n L^{n y_1} + 
    \sum_{m=1}^{m_{\rm max}} 
    b_m (p - p_c)^m~.
\label{fitchi-1}
\end{equation}
As before, we first analyze the data at $\beta = 2$.  To understand the role
of the analytic corrections, we first perform fits of the data in which we fix
$p_c = 0.8925$ and $y_1 = 0.67$, which are the estimates obtained above. If we
do not include the analytic correction (we set $b_m=0$ for any $m$) and we use
$n_{\rm max} = 2$, we obtain $\chi^2/{\rm DOF} = 633/55$, 297/43 from the
analysis of the estimates of $\chi$ corresponding to lattices such that $L\ge
L_{\rm min} = 8$, 12, respectively. If instead we include the analytic
corrections taking $m_{\rm max} = 1$, we obtain $\chi^2/{\rm DOF} = 42/54$,
30/42. The improvement is clearly significant, indicating that the analytic
corrections cannot be neglected.

In Table \ref{res-eta-beta2} we report the results of the fits corresponding
to $n_{\rm max} = 2$ and $m_{\rm max} = 1$. In all cases we fix $y_1$ to
$0.67(2)$, as indicated by Eq.~(\ref{est-y1}).  This is not crucial, since the
estimates of the exponent $\eta$ are quite insensitive to this parameter. The
results show instead a significant dependence on $p_c$ and thus, we present
fits in which $p_c$ is fixed to the value (\ref{est-pc-beta2}) and fits in
which $p_c$ is a free parameter. The estimates of the two fits are
substantially consistent and show a tiny dependence on $L_{\rm min}$. Also the
estimates of $p_c$ are consistent with the value (\ref{est-pc-beta2}).

\begin{table}
  \begin{tabular}{rccccc}
\hline \hline
&\multicolumn{2}{c}{$p_c = 0.8925(1)$} &
\multicolumn{3}{c}{$p_c$ free parameter} \\
\hline
\multicolumn{1}{c}{$L_{\rm min}$}&
\multicolumn{1}{c}{$\chi^2/$DOF}&
\multicolumn{1}{c}{$\eta$} &
\multicolumn{1}{c}{$\chi^2/$DOF}&
\multicolumn{1}{c}{$\eta$} &
\multicolumn{1}{c}{$p_c$} \\
\hline
     8 &  42/54 &   0.1235(13) & 41/53   &  0.1236(4) & 0.89249(2) \\
    12 &  30/42 &   0.1235(15) & 28/41   &  0.1241(5) & 0.89245(4) \\
    16 &  21/30 &   0.1234(16) & 20/29   &  0.1243(8) & 0.89244(5) \\
    24 &  15/19 &   0.1233(20) & 14/18   &  0.1254(20)& 0.89238(11)\\
    32 &   8/7  &   0.1232(23) &  7/6    &  0.1278(40)& 0.89227(20) \\
\hline \hline
  \end{tabular}
  \caption{ Estimates of $\eta$ from fits to Eq.~(\ref{fitchi-1})
   with $n_{\rm max} =2$ and $m_{\rm max} = 1$.
   We fix $y_1 = 0.67(2)$ in both fits.
   The reported error takes into account the error bar on $y_1$ 
   and on $p_c$ (for the fit in which this quantity is fixed). 
   Analyses of the data at $\beta = 2$.}
  \label{res-eta-beta2}
\end{table}

We also considered nonanalytic scaling corrections, performing a fit of the form
\begin{equation}
\ln \chi = (2-\eta) \ln L + \sum_{n=0}^{n_{\rm max}} 
    a_n (p - p_c)^n L^{n y_1} + 
    \sum_{m=1}^{m_{\rm max}} 
    b_m (p - p_c)^m~ + cL^{-\omega}.
\label{fitchi-om}
\end{equation}
We fix $p_c = 0.8925$, $y_1 = 0.67$, $\omega = 0.6$, $n_{\rm max} = 2$,
$m_{\rm max} = 1$, and obtain 
$\eta = 0.1234(13)$, $c = 0.000(3)$ for $L_{\rm min} = 8$: 
there is no evidence of nonanalytic scaling corrections.

To avoid the use of $p_c$, note that Eq.~(\ref{FSS-R})  can be inverted to 
give $u_1 L^{y_1} \approx (p-p_c)L^{y_1}$ as a function of $R$. Thus,
Eq.~(\ref{FSS-chi}) can also be rewritten as 
\begin{equation}
\chi = \bar{u}_h^2 L^{2-\eta} g_\chi(R)[1  + O(L^{-\omega})],
\label{FSS-chi2}
\end{equation}
where $R$ is a renormalized coupling. A polynomial approximation for 
$\bar{u}_h(p)$ and $g_\chi(R)$
gives the fitting form
\begin{equation}
\ln \chi = (2-\eta) \ln L + \sum_{n=0}^{n_{\rm max}} a_n R^n + 
    \sum_{m=1}^{m_{\rm max}} b_m p^m~.
\end{equation}
Fits to this form have a quite large $\chi^2$, which is not unexpected since we
already found that the renormalized couplings show significant scaling
corrections. Moreover, the results depend significantly on the minimum
lattice size $L_{\rm min}$ of the data included in the fit. 
Scaling corrections must therefore be included. We thus consider
\begin{equation}
\ln \chi = (2-\eta) \ln L + \sum_{n=0}^{n_{\rm max}} a_n R^n + 
    \sum_{m=1}^{m_{\rm max}} b_m p^m~ + 
    L^{-\omega} \sum_{k=0}^{k_{\rm max}} c_k R^k.
\label{fitchi-2}
\end{equation}
The results of these fits are reported in Table ~\ref{res-eta-beta2-2}. 
The $\chi^2$ is good; moreover, the results do not depend on which quantity is 
used in the fit, are stable with $L_{\rm min}$, and
are consistent with those reported in 
Table~\ref{res-eta-beta2}.

\begin{table}
  \begin{tabular}{rcccc}
\hline\hline
& \multicolumn{2}{c}{$U_4$} &
\multicolumn{2}{c}{$R_\xi$ }\\
\hline
\multicolumn{1}{c}{$L_{\rm min}$}&
\multicolumn{1}{c}{$\chi^2/$DOF}&
\multicolumn{1}{c}{$\eta$} &
\multicolumn{1}{c}{$\chi^2/$DOF}&
\multicolumn{1}{c}{$\eta$} \\
\hline
     8 &  72/52&   0.1247(5) & 47/52   &  0.1245(14)  \\
    12 &  57/40&   0.1250(8) & 40/40   &  0.1253(22)  \\
    16 &  54/28&   0.1255(9) & 36/28   &  0.1255(48)  \\
\hline
\hline
  \end{tabular}
  \caption{ Estimates of $\eta$ from fits to Eq.~(\ref{fitchi-2})
   with $n_{\rm max} =2$, $m_{\rm max} = 1$, $k_{\rm max} = 0$, 
   $\omega$ free parameter.
   On the left we use $R = U_4$, on the right we use $R = R_\xi$.
   Analyses of the data at $\beta = 2$. }
  \label{res-eta-beta2-2}
\end{table}

\begin{table}
  \begin{tabular}{rccccc}
\hline\hline
&\multicolumn{2}{c}{$p_c = 0.8915(2)$} &
\multicolumn{3}{c}{$p_c$ free parameter} \\
\hline
\multicolumn{1}{c}{$L_{\rm min}$}&
\multicolumn{1}{c}{$\chi^2/$DOF}&
\multicolumn{1}{c}{$\eta$} &
\multicolumn{1}{c}{$\chi^2/$DOF}&
\multicolumn{1}{c}{$\eta$} &
\multicolumn{1}{c}{$p_c$} \\
\hline
      8 &  25/28 & 0.1336(26)  & 23/27 & 0.1342(7) & 0.89145(3) \\
     12 &  23/23 & 0.1335(30)  & 22/22 & 0.1341(9) & 0.89145(4) \\
     16 &  21/18 & 0.1335(33)  & 20/17 & 0.1341(12)& 0.89146(6) \\
     24 &  16/13 & 0.1330(38)  & 16/12 & 0.1315(23)& 0.89158(12) \\
     32 &  10/8  & 0.1330(42)  & 10/7  & 0.1361(46)& 0.89133(21) \\
\hline
\hline
  \end{tabular}
  \caption{ Estimates of $\eta$ from fits to Eq.~(\ref{fitchi-1})
   with $n_{\rm max} =2$ and $m_{\rm max} = 1$.
   We fix $y_1 = 0.67(2)$ in both fits.
   The reported error takes into account the error bar on $y_1$ 
   and on $p_c$ (for the fit in which this quantity is fixed). 
   Analyses of the data at $\beta = 1.55$.}
  \label{res-eta-beta1.55}
\end{table}

Analogous analyses can be performed at $\beta = 1.55$. Also in this case the
analytic corrections cannot be neglected and thus we only consider fits with
$m_{\rm max} = 1$. The results of the fits to Eq.~(\ref{fitchi-1}) are
reported in Table~\ref{res-eta-beta1.55}. The dependence of the results on
$L_{\rm min}$ is tiny. Moreover, the estimates of $p_c$ obtained in the
analyses in which this quantity is a free parameter are perfectly consistent
with the estimate (\ref{est-pc-beta1.55}).  Similar, though less stable,
results are obtained by fitting the data to Eq.~(\ref{fitchi-2}). We fix $\omega = 0.6(1)$ as in the case of the analyses of the renormalized couplings.
For $L_{\rm min} = 8$ we obtain $\eta = 0.1304(5)$ and $\eta =
0.1340(5)$ by using $U_4$ and $R_\xi$, respectively; for $L_{\rm min} = 12$ we
obtain instead $\eta = 0.1317(8)$ and $\eta = 0.1329(8)$ .

Collecting all results, from the analyses of the data at $\beta = 2$ 
we would estimate $\eta = 0.125(3)$. The analyses at $\beta =
1.55$ give a slightly different value, $\eta = 0.132(4)$. The difference is
tiny---less than two combined error bars---but indicates that there are
corrections which are not fully taken into account by our analyses.  As final
estimate we report the average of the two results,
\begin{equation}
\eta = 0.128(8).
\label{etaes}
\end{equation}
The error we quote is quite conservative and essentially includes the 
estimates of all fits for both values of $\beta$.

\subsection{The exponent $\beta$ and a  
check of hyperscaling} 

The exponent $\beta$ can be determined from the critical behavior of the 
magnetization. The RG predicts 
\begin{equation}
m = \overline{u}_h L^{-\beta/\nu} f_m[(p-p_c) L^{y_1}],
\end{equation}
where $\bar{u}_h$ is the same function which appears in 
Eq.~(\ref{FSS-chi}) and $f_m(x)$ is a universal function.
Expanding this scaling relation around the critical point
we obtain the fitting form
\begin{equation}
\ln m = - {\beta \over \nu} \ln L + \sum_{n=0}^{n_{\rm max}}
    a_n (p - p_c)^n L^{n y_1} +
    \sum_{m=1}^{m_{\rm max}}
    b_m (p - p_c)^m~.
\label{fitm-1}
\end{equation}
As before we fix $n_{\rm max} = 2$, $m_{\rm max} = 1$, and use 
the best available estimates of $p_c$. For $\beta = 2$ a fit of the data
satisfying $L\ge L_{\rm min} = 24$ gives $\beta/\nu = 0.0613(11)$; 
for $L_{\rm min} = 32$ we obtain instead $\beta/\nu = 0.0614(12)$. 
For $\beta = 1.55$ and $L_{\rm min} = 32$ we obtain 
$\beta/\nu = 0.0661(22)$. As in the case of $\eta$, we observe a tiny 
difference between the estimates obtained at the two values of the temperature.
It probably indicates the presence of additional scaling corrections 
which are not taken into account by our scaling Ansatz. A 
conservative estimate of the critical exponent which is consistent 
with all results is 
\begin{equation}
{\beta \over \nu} = 0.063(3) \qquad\qquad 
   \beta = 0.095(5).
\end{equation}
We can now check hyperscaling. If it holds, we should have 
$2\beta/\nu - \eta + 2 = d = 2$. We find
\begin{equation}
{2\beta \over \nu} - \eta + 2 = 2.00(1).
\end{equation}
Hyperscaling is verified quite precisely. 

\begin{figure*}[tb]
\centerline{\psfig{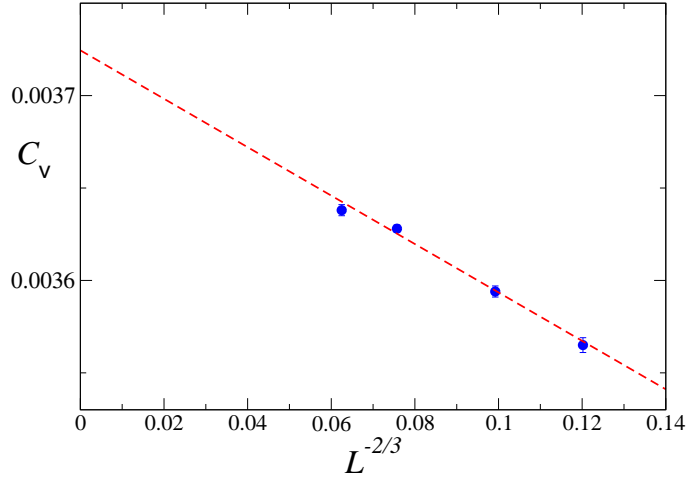}} 
\caption{The specific heat at $\beta = 2$ and $p = 0.8925\approx p_c$
versus $L^{-2/3}$. The dashed line is obtained by fitting the MC data 
to $a + b L^{-2/3}$.
}
\label{Cv}
\end{figure*}

Finally, we consider the specific heat. At $p = p_c$ we expect
\begin{equation}
C_v = a + b L^{\alpha/\nu},
\label{specheat-scal}
\end{equation}
where $a$ is due to the analytic contribution to the free energy. 
If hyperscaling holds, we should have $\alpha = 2 - 2\nu$, so that 
\begin{equation}
{\alpha\over \nu} = {2\over \nu} - 2 = 2 y_1 - 2 = -0.66(4).
\end{equation}
A precise determination of $\alpha/\nu$ from the data is quite difficult,
because $\alpha/\nu < 0$---the singular part decreases as $L\to \infty$.
Thus, we have only checked that our data are consistent with hyperscaling.
In Fig.~\ref{Cv} we show the specific heat for $\beta = 2$ and $L\ge 24$
versus $L^{-2/3}$. The results are consistent, supporting
hyperscaling.

\subsection{The derivative $dp_c/d\beta$}

As a final test of our results we consider the derivative with 
respect to $\beta$ of a renormalized coupling $R$. In the FSS limit 
$R$ behaves as 
\begin{equation}
R = f_R(u_1 L^{y_1}) + u_\omega L^{-\omega} f_{R,\omega}(u_1 L^{y_1}) + 
\label{FSSR-2}
\ldots
\end{equation}
where the scaling fields $u_1$ and $u_\omega$ are functions of the system
parameters, hence of $\beta$ and $p$. Moreover, $u_1$ vanishes on the critical
line. From Eq.~(\ref{FSSR-2}) we obtain
\begin{equation}
{\partial R\over \partial\beta} = 
    {\partial u_1\over \partial\beta} L^{y_1} 
    \left[ f_R'(u_1 L^{y_1}) + 
     u_\omega L^{-\omega} f_{R,\omega}'(u_1 L^{y_1})\right] + 
     {\partial u_\omega\over \partial\beta} 
     L^{-\omega} f_{R,\omega}(u_1 L^{y_1}) + \ldots
\label{derR}
\end{equation}
If the critical value $p_c$ is $\beta$ independent, ${\partial
  u_1/ \partial\beta}$ vanishes on the critical line, so that ${\partial
  R/ \partial\beta}$ behaves as $L^{-\omega}$ for $L\to\infty$, i.e. the
derivative vanishes in the critical large-$L$ limit.  This is not surprising,
since for $p = p_c = p^*$ and any $\beta$ we would have $R = R^* +
O(L^{-\omega})$, with $R^*$ independent of $\beta$.  On the other hand, if the
transition is reentrant, ${\partial R/ \partial\beta}$ diverges as $L^{y_1}$.

We have checked the validity of Eq.~(\ref{derR}) by using the data 
at $\beta = 2$. The fits of the renormalized 
couplings $R$ give us estimates of the expansion of $R$ around $p_c$. 
In particular, fits to Eq.~(\ref{fitR-2}) give us estimates of the 
coefficients $a_n$. We have thus fitted $\partial R/\partial\beta$ 
to the following expression:
\begin{equation}
{\partial R\over \partial\beta} = k_0 L^{y_1} 
   [a_1 + 2 a_2 (p - p_c) L^{y_1}] + k_1 L^{y_1 - \omega}.
\label{fit-derR}
\end{equation}
We take $y_1 = 0.67(2)$, $p_c = 0.8925(1)$, $\omega = 0.6(1)$,
and $a_1$ and $a_2$ from the fits of $R$ to Eq.~(\ref{fitR-2});
$k_0$ and $k_1$ are free parameters.
The estimates of $k_0$  do not vary significantly with $L_{\rm min}$. 
Moreover, results obtained by using 
${\partial R_\xi/ \partial\beta}$ and
${\partial U_4/ \partial\beta}$ are fully consistent.
Comparing all results we obtain the estimate
\begin{equation}
k_0 = -0.0020(3)~.
\end{equation}
To interpret this result, note that Eqs.~(\ref{FSS-R}), (\ref{exp-u1}), and 
(\ref{fitR-2}) allow us to identify 
\begin{equation}
a_1 = A_0(\beta) f_R'(0).
\end{equation}
Instead, comparing Eq.~(\ref{fit-derR}) with Eq.~(\ref{derR}) we obtain 
\begin{equation}
k_0 a_1 = \left. {\partial u_1\over \partial \beta}\right|_{p_c} f_R'(0).
\end{equation}
Now, Eq.~(\ref{exp-u1}) gives 
\begin{equation}
  \left. {\partial u_1\over \partial \beta}\right|_{p_c} = 
   - A_0(\beta) {dp_c\over d\beta}.
\end{equation}
It follows 
\begin{equation} 
  {dp_c\over d\beta} = -k_0 = 0.0020(3)~.
\end{equation}
Again, this result shows that the transition is reentrant. It is also
consistent with the crude estimate 
\begin{equation}
{dp_c\over d\beta} \approx 
    {p_c(2) - p_c(1.55)\over 2 - 1.55} \approx 0.0022.
\end{equation}

\section{Conclusions}
\label{conclusions}

In this paper we have studied the nature of the transition line which starts
from the MNP and ends at $T = 0$ and which separates the paramagnetic phase
from the ferromagnetic phase.  For this purpose, we have presented FSS
analyses of MC data on lattices of linear size $L$ up to $L=64$ for
$\beta\equiv 1/T=2$ and $\beta=1.55$.

Our main results are the following.
\begin{itemize}
\item[(i)] The PF transition line below the MNP is reentrant. 
Indeed, we find $p_c=0.8915(2)$ at
$T=1/1.55\approx 0.645$ and $p_c=0.8925(1)$ at $T=0.5$. Therefore,
$p_c>p^*=0.89083(3)$ for any $T<T^*=0.9527(1)$, where $X^*=(T^*,p^*)$ is the
location of the MNP.  
  
\item[(ii)] The PF transitions are of second order with a
  standard power-law behavior.

\item[(iii)] The estimated values of the critical exponents and of the
  large-$L$ limit of the RG invariant quantities 
  $U_4$, $R_\xi$, and $U_{22}$ at two different points of
  the line ($\beta = 1.55$ and $\beta = 2$) suggest that the PF
  transitions for $0<T<T^*$
  belong to a unique universality class. In particular, 
the corresponding critical exponents are
\begin{equation}
\nu = 1.50(4),\qquad \eta = 0.128(8), \qquad \beta = 0.095(5)~.
\label{estimate-fin-c}
\end{equation}
They satisfy the hyperscaling relation $2\beta/\nu - \eta = d - 2 = 0$.  Our
MC data are also consistent with the hyperscaling relation $\alpha = 2 - d\nu
= 2 - 2\nu$, which gives $\alpha = -1.00(8)$.  Using the scaling relation
$\gamma = (2 - \eta)\nu$, we derive $\gamma = 2.81(8)$.  The estimates
(\ref{estimate-fin-c}) are definitely different from the Ising values $\nu=1$,
$\eta=1/4$, $\beta = 1/8$.  We note that they are consistent with the simple
rational expressions $\nu = 3/2$, $\eta = 1/8$.

\item[(iv)] The above results show that 
  in two dimensions there are two
  fixed points which control the PF transitions in
  disordered random-bond Ising systems: besides the standard Ising fixed point,
  which is relevant for small disorder and controls the critical behavior
  along the PF transition line for $T^*<T\le T_{\rm Is}$, there is also a
  strong-disorder fixed point which controls the critical behavior along the
  PF transition line for $0<T<T^*$. 
  The resulting phase diagram is consistent with the results of 
  Refs.~\cite{McMillan-84,MB-98}.
Note that frustration and not simply disorder is
the relevant property, which gives rise to the new fixed point.  Indeed, in
randomly-dilute Ising systems, in which there is dilution but not frustration,
there is no evidence of a new strong-disorder fixed point \cite{HPPV-08-2}.
\end{itemize}

\begin{figure*}[tb]
\centerline{\psfig{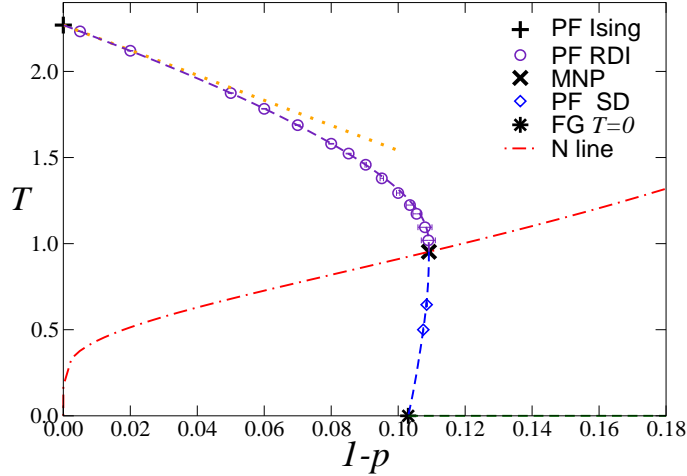}}
\caption{
The phase diagram of the square-lattice $\pm J$ Ising model.
The estimates of the critical points for $T>T^*$ are taken from 
Refs.~\protect\cite{MC-02,HPPV-08-2}. The estimate of the
$T=0$ transition point is taken from Refs.~\cite{AH-04,WHP-03}. 
The dashed lines 
are interpolations discussed in App.~\ref{MNP}, while 
the dotted line starting at the Ising point corresponds to the 
approximation (\ref{TCder}). 
}
\label{phadiatc}
\end{figure*}

It is interesting to compare our results with those obtained at $T=0$.
Ref.~\cite{McMillan-84} extrapolated the RG results to $T=0$ (this is correct
under the assumption that the limit $T\to 0$ is regular) and obtained $\nu =
1.42(8)$.  Ref.~\cite{WHP-03} obtained $\nu=1.46(1)$ from the scaling of the
failure probability.  Ref.~\cite{AH-04} found $\nu=1.55(1)$ from the analysis
of the Binder cumulant and the magnetization exponent
$\beta=0.09(1)$.\footnote{Ref.~\cite{AH-04} reports $\beta = 0.9(1)$.
  Alexander Hartmann communicated to us that the correct result is $\beta =
  0.09(1)$.}  They also analyzed the domain-wall energy, obtaining $\Delta E =
L^\rho f\left( (p-p_0)L^{y_1}\right)$ with $\rho=0.12(5)$ and $y_1=0.75(5)$.
This gives $\nu=1/y_1 = 1.33(9)$.  The exponent $\eta$ associated with the
spin-spin correlation has been estimated in Ref.~\cite{PHP-06}, obtaining
$\eta\approx 0.13$. Note that these estimates are consistent with the
hyperscaling relation $2\beta/\nu - \eta = d - 2 = 0$.  Therefore, even if
these results refer to a $T=0$ transition, the magnetic exponents are
consistent with hyperscaling.

The $T=0$ results are very close to ours.  Note also that, at $T=0$, the
relation $U_4^* = U_{22}^*+1$ holds.\footnote{Indeed, assuming a nondegenerate
  ground state (this should be the case in two dimensions), we have $\mu_4 =
  \mu_2^2$ (see App.~\ref{notations} for the definitions), from which the
  relation follows.} This relation is approximately satisfied by our
finite-$T$ data, see Eqs.~(\ref{U4-beta2}) and (\ref{U22-beta2}); the slight
discrepancy might be due to the presence of neglected additional scaling
corrections.  All results are therefore consistent with a single magnetic
fixed point that controls the magnetic critical behavior both at $T > 0$ and
at $T=0$.  At the multicritical $T=0$ point, glassy and magnetic modes are
apparently effectively decoupled.

Finally, we have improved the estimates of the critical parameters at the MNP,
by a new FSS analysis of MC simulations up to $L=64$ along the N line.  We
obtain $p^*=0.89083(3)$ and $T^*=0.9527(1)$,
$y_1=0.66(1)$ and $y_2=0.250(2)$ for the RG dimensions of the two relevant
operators in the absence of external field, and $\eta=0.177(2)$ for the
magnetic critical exponent associated with the spin-spin correlation function.

In Fig.~\ref{phadiatc} we report the available estimates of the critical
points and report simple interpolations, discussed in App.~\ref{MNP}, which
take into account all theoretical predictions and numerical results.

\bigskip
Discussions with Marco Picco and correspondence with Nihat Berker and
Alexander Hartmann are gratefully acknowledged.

\appendix

\section{Some details on the Monte Carlo simulations} 
\label{MC-details}

\begin{table}
  \begin{tabular}{clllccrr}
\hline
\hline
\multicolumn{1}{c}{$L$}&
\multicolumn{1}{c}{$p$}&
\multicolumn{1}{c}{$\beta_{\rm min}$}&
\multicolumn{1}{c}{$\beta_{\rm max}$}&
\multicolumn{1}{c}{$N_T$}&
\multicolumn{1}{c}{$N_{\rm s}/64$}&
\multicolumn{1}{c}{$N_{\rm run}/10^3$}&
\multicolumn{1}{c}{$N_{\rm therm}/10^3$}\\
\hline
$32$ & $0.8910$ & $0.800$ & $1.55$ & $5$ & $15625$ & $800$ & $240$ \\
$32$ & $0.8915$ & $0.800$ & $1.55$ & $5$ & $15625$ & $800$ & $240$ \\
$32$ & $0.8920$ & $0.800$ & $1.55$ & $5$ & $15625$ & $800$ & $240$ \\
$32$ & $0.8925$ & $0.800$ & $1.55$ & $5$ & $15625$ & $800$ & $240$ \\
$32$ & $0.8930$ & $0.800$ & $1.55$ & $5$ & $15625$ & $800$ & $400$ \\
$32$ & $0.8915$ & $0.800$ & $2$ & $6$ & $15625$ & $800$ & $240$ \\
$32$ & $0.8920$ & $0.800$ & $2$ & $6$ & $15625$ & $800$ & $240$ \\
$32$ & $0.8925$ & $0.800$ & $2$ & $6$ & $15625$ & $800$ & $240$ \\
$32$ & $0.8930$ & $0.800$ & $2$ & $6$ & $15625$ & $800$ & $240$ \\
$48$ & $0.8915$ & $0.740$ & $2$ & $9$ & $31250$ & $2000$ & $400$ \\
$48$ & $0.8920$ & $0.740$ & $2$ & $9$ & $31250$ & $2000$ & $400$ \\
$48$ & $0.8925$ & $0.740$ & $2$ & $9$ & $31250$ & $2000$ & $400$ \\
$48$ & $0.8930$ & $0.740$ & $2$ & $9$ & $31250$ & $2000$ & $600$ \\
$64$ & $0.8915$ & $0.710$ & $2$ & $13$& $7813$  & $3000$ & $900$ \\
$64$ & $0.8920$ & $0.710$ & $2$ & $13$& $7813$ & $3000$ & $900$ \\
$64$ & $0.8925$ & $0.710$ & $2$ & $13$& $7813$ & $3000$ & $900$ \\
$64$ & $0.8930$ & $0.710$ & $2$ & $13$& $7813$ & $3000$ & $900$ \\
\hline\hline
\end{tabular}
\caption{Parameters of the random-exchange MC runs for $L \ge 32$.}
  \label{mcruns}
\end{table}

In our parallel-tempering simulations we consider $N_T$ systems at the same
value of $p$ and at $N_T$ different inverse temperatures $\beta_{\rm min}
\equiv \beta_1$, \ldots, $\beta_{N_T} \equiv \beta_{\rm max}$, where 
$\beta_{\rm max}$ is chosen to be either 2 or 1.55. To avoid repeating the 
runs twice, for $L = 48$ and 64, $\beta_{\rm max}$ is always chosen to be 2, 
while one of the $\beta_i$ corresponds to 1.55.
Moreover, for all values of $L$, we choose $\beta_i = \beta_N(p)$ for some 
$i$, where $\beta_N(p)$ is given in Eq.~(\ref{tn}), so that the corresponding 
point lies on the N line.  This choice gives us estimates along the N line, 
which can be compared with exact and previous numerical 
results.  They provide a check of the numerical simulations and allow us 
to improve the estimates of the critical parameters of Ref.~\cite{HPPV-08},
see App.~\ref{MNP}.

The elementary unit of the algorithm consists in $N_{\rm ex}=20$
Metropolis sweeps for each configuration followed by an exchange move.  We
consider all pairs of configurations corresponding to nearby temperatures
and propose a temperature exchange with acceptance probability
\begin{equation}
  {\cal P} = \exp \{(\beta_i - \beta_{i+i})(E_i-E_{i+1})\},
\end{equation}
where $E_i$ is the energy of the system at inverse temperature $\beta_i$.  We
generate $N_s$ disorder samples, and for every sample we perform a MC run of
$N_{\rm run}$ Metropolis sweeps for each $\beta_i$ value.  The first $N_{\rm
  therm}$ iterations are discarded for thermalization (see Ref.~\cite{HPPV-08}
for a discussion of the thermalization issues). The parameters of the runs
with $L\ge 32$ are reported in Table~\ref{mcruns}. Finally, note that the
determination of $U_{22}$ requires the computation of a disorder average of
the square of a thermal average. We use an essentially bias-free estimator
discussed in Ref.~\cite{HPPV-07}.

\section{Definitions} \label{notations}

The two-point correlation function is defined as
\begin{equation}
G(x) \equiv [ \langle \sigma_0 \,\sigma_x \rangle ],
\label{twof}
\end{equation}
where the angular and the square brackets indicate the thermal
average and the quenched average over disorder, respectively. 
We define the magnetic
susceptibility $\chi\equiv \sum_x G(x)$ and the correlation length $\xi$,
\begin{equation}
\xi^2 \equiv {\widetilde{G}(0) - \widetilde{G}(q_{\rm min}) \over 
          \hat{q}_{\rm min}^2 \widetilde{G}(q_{\rm min}) },
\end{equation}
where $q_{\rm min} \equiv (2\pi/L,0)$, $\hat{q} \equiv 2 \sin q/2$, and
$\widetilde{G}(q)$ is the Fourier transform of $G(x)$.  We also 
consider the magnetization $m$ defined as 
\[
m = {1\over V} \left[ \langle | \sum_x \sigma_x |
        \rangle \right],
\]
where $V$ is the volume, and the specific heat $C_v$
\[
C_v = {1\over V} \left[ 
    \langle {\cal H}^2 \rangle - \langle {\cal H} \rangle^2 \right],
\]
where $ {\cal H}$ is the Hamiltonian.

We also consider
quantities (we call them renormalized couplings) that are invariant under RG
transformations in the critical limit.  Beside the ratio
\begin{equation}
R_\xi \equiv \xi/L,
\label{rxi}
\end{equation}
we consider the RG invariant quantities 
\begin{eqnarray}
U_{4}  \equiv { [ \mu_4 ]\over [\mu_2]^{2}}, \qquad
U_{22} \equiv  {[ \mu_2^2 ]-[\mu_2]^2 \over [\mu_2]^2},\qquad
U_d \equiv  U_4 - U_{22},
\nonumber 
\end{eqnarray}
where
\begin{eqnarray}
\mu_{k} \equiv \langle \; ( \sum_x \sigma_x\; )^k \rangle.
\end{eqnarray}

\section{Critical exponents at the multicritical point} \label{MNP}

In each parallel-tempering simulation we fixed $p$ and considered several
values of $\beta$ from $\beta_{\rm min} < \beta^*$ up to $\beta_{\rm max}$
which is either 2 or 1.55, hence larger than the multicritical value
$\beta^*$. In all runs we were careful to include a point on the N
line. Since the energy is known exactly on this line, this choice allowed us
to test the correctness of the simulation code. Moreover, we were
able to collect a significant amount of new data, which can be combined with
the old ones presented in Ref.~\cite{HPPV-08}.  As we shall see, the FSS
analyses of this new set of data allows us to improve the estimates of 
the critical parameters.  

\begin{table}
  \begin{tabular}{clcll}
\hline
\hline
&
\multicolumn{1}{c}{$L_{\rm min}$}&
\multicolumn{1}{c}{$\chi^2$/DOF}&
\multicolumn{1}{c}{$p^*$}&
\multicolumn{1}{c}{$y_1$}\\
\hline
$R_\xi$,$U_4$,$U_{22}$
         &  12      &        383/289   &      0.890864(4)  &   0.659(2)\\
         &  16      &        207/220   &      0.890844(4)  &   0.658(2)\\
         &  24      &        120/151   &      0.890828(6)  &   0.658(3)\\
         &  32      &         58/82    &      0.890822(8)  &   0.651(5)\\
$R_\xi$,$U_4$,$U_{d}$
         &  12      &        424/289   &      0.890853(3)  &   0.660(1) \\
         &  16      &        248/220   &      0.890856(4)  &   0.660(2) \\
         &  24      &        194/151   &      0.890850(5)  &   0.659(5) \\
         &  32      &        100/82    &      0.890848(7)  &   0.653(5)\\
\hline
\hline
  \end{tabular}
  \caption{Estimates of $p^*$ and $y_1$ at the MNP. Results from combined 
   fits of three different renormalized couplings to Eq.~(\ref{fitR-1})
   with $n_{\rm max} = 2$. Here $U_d \equiv  U_4 - U_{22}$. }
  \label{MNP-table1}
\end{table}

As in Ref.~\cite{HPPV-08} we perform combined fits of the renormalized 
couplings
to Eq.~(\ref{fitR-1}) and (\ref{fitR-2}). The new results are reported in 
Tables~\ref{MNP-table1} and \ref{MNP-table2}. The estimates of $y_1$
are quite stable and essentially independent of $L_{\rm min}$, 
of the observable, and of the scaling corrections. We thus quote
\begin{equation}
    y_1 = 0.66(1),
\end{equation}
where the error is chosen quite conservatively, and is such to include
all results. This result is fully consistent 
with the estimate $y_1 = 0.655(15)$ of Ref.~\cite{HPPV-08}.
The estimates of $p^*$ vary between 0.89081 and 0.89086, so that we quote
\begin{equation}
    p^* = 0.89083(3).
\label{pstar-MNP}
\end{equation}
This estimate agrees with that we obtained in Ref.~\cite{HPPV-08},
i.e. $p^*=0.89081(7)$.  Moreover, it is in full agreement with the recent
calculations of Ref.~\cite{Ohzeki-08}: Two different approximations gave
$p^*\approx 0.890822$ and $p^*\approx 0.890813$.

Our analyses also provide estimates of the critical-point value of the 
renormalized couplings:
\begin{eqnarray}
R_\xi^* &=& 0.997(1), \\
U_4^*   &=& 1.1264(4), \\
U_{22}^*&=& 0.0817(3).
\end{eqnarray}
Scaling corrections are particularly weak and 
apparently decay as $L^{-2}$ or faster. Note that this does not necessarily
imply the presence of nonanalytic corrections associated with 
RG irrelevant operators with $\omega \approx 2$. Indeed, in all cases
we expect contributions due to the regular part of the free energy, 
which decay as $L^{\eta - 2} \approx L^{-1.8}$. 

\begin{table}
  \begin{tabular}{clclll}
\hline
\hline
&
\multicolumn{1}{c}{$L_{\rm min}$}&
\multicolumn{1}{c}{$\chi^2$/DOF}&
\multicolumn{1}{c}{$p^*$}&
\multicolumn{1}{c}{$y_1$}&
\multicolumn{1}{c}{$\omega$} \\
\hline
$R_\xi$,$U_4$,$U_{22}$
          &  6  & 283/318 &  0.890822(7) &    0.665(3)  &  1.79(13) \\
          &  8  & 232/300 &  0.890814(9) &    0.660(10) &  1.98(25) \\
$R_\xi$,$U_4$,$U_{d}$
          &  6  & 396/318 &  0.890864(3)&    0.662(2)  &  3.18(10) \\
          &  8  & 332/300 &  0.890857(4)&    0.660(2)  &  4.41(27) \\
\hline
\hline
  \end{tabular}
  \caption{Estimates of $p^*$, $y_1$, and $\omega$ at the MNP. 
   Results from combined 
   fits of three different renormalized couplings to Eq.~(\ref{fitR-2})
   with $n_{\rm max} = 2$ and $k_{\rm max} = 1$. }
  \label{MNP-table2}
\end{table}

\begin{table}
  \begin{tabular}{clclcl}
\hline
\hline
&
\multicolumn{1}{c}{$L_{\rm min}$}&
\multicolumn{1}{c}{$\chi^2$/DOF}&
\multicolumn{1}{c}{$y_2$}&
\multicolumn{1}{c}{$\chi^2$/DOF}&
\multicolumn{1}{c}{$y_2$} \\
\hline
$R_\xi'$
  &   8  &    90/102&  0.2533(6)[5] & 57/101 & 0.2521(5)[5] \\
  &  12  &    88/96 &  0.2535(7)[5] & 56/95  & 0.2519(6)[6] \\
  &  16  &    60/73 &  0.2530(9)[6] & 45/72  & 0.2514(8)[7] \\
  &  24  &    37/50 &  0.2531(13)[7]& 31/49  & 0.2515(13)[8]\\
  &  32  &    16/27 &  0.2528(21)[8]& 15/26  & 0.2523(21)[8] \\
$U_4'$
  &    8 &    158/102&  0.2492(9)[18]  & 90/101& 0.2480(4)[17] \\ 
  &   12 &    148/96 &  0.2496(11)[19] & 89/95 & 0.2478(5)[18] \\
  &   16 &     95/73 &  0.2450(14)[21] & 65/72 & 0.2480(7)[21] \\
  &   24 &     53/50 &  0.2509(18)[25] & 42/49 & 0.2490(11)[24]\\
  &   32 &     32/27 &  0.2514(23)[28] & 29/26 & 0.2500(19)[27]\\
\hline
\hline
  \end{tabular}
  \caption{Estimates of $y_2$ at the MNP. We fix $y_1 = 0.66(1)$ and 
   $p_c = 0.89083(3)$.
   Results from fits of the derivative of the renormalized couplings 
   $R_\xi$ and $U_4$. On the left 
   the results refer to the fit to  Eq.~(\ref{fitRp-1}) 
   with $n_{\rm max} = 2$,  
   on the right to the fit to Eq.~(\ref{fitRp-2}) with 
   $n_{\rm max} = 2$ and $m_{\rm max} = 1$.
   The error in parentheses is the sum of the statistical error and of the 
   error due to uncertainty of $y_1$; the error in brackets gives the 
   variation of the estimate as $p_c$ varies by one error bar. }
  \label{MNP-table3}
\end{table}

The critical exponent $y_2$ is derived from the critical behavior 
of $R'\equiv \partial R/\partial \beta$, where $R$ is 
a renormalized coupling \cite{HPPV-08}. Neglecting scaling 
correction, its FSS behavior is given by
\begin{equation}
R' = {\partial u_1\over\partial \beta} L^{y_1} 
    f_1(u_1 L^{y_1},u_2 L^{y_2}) + 
     {\partial u_2\over\partial \beta} L^{y_2}
    f_2(u_1 L^{y_1},u_2 L^{y_2}),
\label{Rp-MNP}
\end{equation}
where $u_1$ and $u_2$ are the nonlinear scaling fields associated with the 
two leading  relevant operators. In general, we expect \cite{LH-89}
$u_2$ to vanish on the N line, so that 
\begin{equation}
u_2(\beta,p) = S(\beta - \beta_N(p),p-p^*),
\label{def-u2-MNP}
\end{equation}
where $\beta_N(p) = 1/T_N(p)$, $T_N(p)$ is defined in Eq.~(\ref{tn}), and
 the function $S(x,y)$ is such that $S(0,y) = 0$ and 
$\partial S(0,0)/\partial x\not=0$. 
Since the transition lines must be tangent to the 
line $p = p^*$ as a consequence of a general rigorous inequality
\cite{Nishimori-81}, we also have 
\begin{equation}
u_1(\beta,p) = p - p^* + \hbox{\rm quadratic terms}.
\label{u1def}
\end{equation}
The independence of $u_1$ on $\beta$ at leading order, implies that the 
first term in Eq.~(\ref{Rp-MNP}) vanishes at the MNP, so that 
$R'\sim L^{y_2}$ for $L\to \infty$ at $p = p^*$. 

In order to compute $y_2$, we perform three different fits of our 
data on the N line. In the first
one, we neglect the $p$ dependence of $\partial u_2/\partial\beta$ and 
set $\partial u_1/\partial\beta = 0$. Then, setting $u_2 = 0$ and 
expanding in powers of $u_1 L^{y_1} \sim (p - p^*) L^{y_1}$, we obtain
\begin{equation}
\ln R' = y_2 \ln L + \sum_{n=0}^{n_{\rm max}} 
    a_n (p - p^*)^n L^{n y_1}.
\label{fitRp-1}
\end{equation}
In the second fit we include the nontrivial dependence of $u_2$
on $\beta$ and $p$.  We fit the results to 
\begin{equation}
\ln R' = y_2 \ln L + \sum_{n=0}^{n_{\rm max}} 
    a_n (p - p^*)^n L^{n y_1} + 
    \sum_{m=1}^{m_{\rm max}} b_m (p-p^*)^m.
\label{fitRp-2}
\end{equation}
Finally, note that $u_1$ may depend on $\beta$ at quadratic and higher orders,
so that on the N line one may have
\begin{equation}
{\partial u_1\over \partial\beta} \sim p - p^* + O[(p - p^*)^2].
\end{equation}
Hence, the first term in Eq.~(\ref{Rp-MNP}) may give rise to 
corrections of order $(p - p^*) L^{y_1 - y_2}$. Thus, we also perform fits to
\begin{equation}
\ln R'
 = y_2 \ln L + \sum_{n=0}^{n_{\rm max}} 
    a_n (p - p^*)^n L^{n y_1} + L^{-y_2} 
       \sum_{k=1}^{k_{\rm max}} b_k (p - p^*)^k L^{k y_1}.
\label{fitRp-3}
\end{equation}
In Table \ref{MNP-table3} we report the results of the fits of $R'_\xi$ and
$U_4'$ to Eqs.~(\ref{fitRp-1}) and (\ref{fitRp-2}).
The inclusion of the analytic corrections significantly reduces the $\chi^2$
and changes slightly the estimates of $y_2$.
Fits to  Eq.~(\ref{fitRp-3}) give results which are essentially equivalent 
to those obtained by fitting to Eq.~(\ref{fitRp-2}).
Comparing all results we obtain the estimate
\begin{equation}
y_2 = 0.250(2),
\end{equation}
which is identical to that reported in Ref.~\cite{HPPV-08}. 

\begin{table}
  \begin{tabular}{lclcl}
\hline
\hline
& \multicolumn{2}{c}{$\ln Z$}&
   \multicolumn{2}{c}{$\ln \chi$} \\
\hline
\multicolumn{1}{c}{$L_{\rm min}$}&
\multicolumn{1}{c}{$\chi^2$/DOF}&
\multicolumn{1}{c}{$\eta$}&
\multicolumn{1}{c}{$\chi^2$/DOF}&
\multicolumn{1}{c}{$\eta$} \\
\hline
      8 &   393/102&  0.1736(6)[12] & 2653/102  & 0.1752(2)[5] \\
     12 &   272/96 &  0.1747(7)[13] & 2340/96   & 0.1749(3)[5] \\
     16 &   146/73 &  0.1760(8)[14] & 1342/73   & 0.1751(3)[5] \\
     24 &    68/50 &  0.1776(10)[16]&  761/50   & 0.1752(4)[6] \\
     32 &    30/27 &  0.1782(12)[18]&  110/27   & 0.1761(4)[7] \\
\hline
\hline
  \end{tabular}
  \caption{Estimates of $\eta$ at the MNP. We fix $y_1 = 0.66(1)$ and 
   $p_c = 0.89083(3)$.
   Results from fits of $\ln Z$ and $\ln \chi$ to Eq.~(\ref{fitchi-1}) with
   $n_{\rm max} = 2$ without analytic correction (in the case of 
   $\ln Z$ the coefficient of $\ln L$ is of course $-\eta$).
   The error in parentheses is the sum of the statistical error and of the 
   error due to uncertainty of $y_1$; the error in brackets gives the 
   variation of the estimate as $p_c$ varies by one error bar. }
  \label{MNP-table4}
\end{table}

\begin{table}
  \begin{tabular}{lclcl}
\hline
\hline
& \multicolumn{2}{c}{$\ln Z$}&
   \multicolumn{2}{c}{$\ln \chi$} \\
\hline
\multicolumn{1}{c}{$L_{\rm min}$}&
\multicolumn{1}{c}{$\chi^2$/DOF}&
\multicolumn{1}{c}{$\eta$}&
\multicolumn{1}{c}{$\chi^2$/DOF}&
\multicolumn{1}{c}{$\eta$} \\
\hline
      8 &   300/101 & 0.1745(3)[11] & 83/101  & 0.1767(1)[5] \\
     12 &   117/95  & 0.1763(3)[12] & 79/95   & 0.1768(1)[6] \\
     16 &   60/72   & 0.1776(4)[14] & 50/72   & 0.1774(1)[6] \\
     24 &   32/49   & 0.1791(5)[15] & 36/49   & 0.1771(1)[7] \\
     32 &   15/26   & 0.1794(9)[17] & 20/26   & 0.1771(2)[7] \\
\hline
\hline
  \end{tabular}
  \caption{Estimates of $\eta$ at the MNP. We fix $y_1 = 0.66(1)$ and 
  $p_c = 0.89083(3)$.
   Results from fits of $\ln Z$ and $\ln \chi$ to Eq.~(\ref{fitchi-1}) with
   $n_{\rm max} = 2$ and $m_{\rm max} = 1$.
   The error in parentheses is the sum of the statistical error and of the 
   error due to uncertainty of $y_1$; the error in brackets gives the 
   variation of the estimate as $p_c$ varies by one error bar. }
  \label{MNP-table5}
\end{table}

Finally, we determine $\eta$. We compute it from the critical 
behavior of $\chi$ and, as in Ref.~\cite{HPPV-08}, from that of 
$Z\equiv \chi/\xi^2$. The results of the fits with and without analytic
corrections are reported in Tables \ref{MNP-table4} and \ref{MNP-table5}.
The most stable results are obtained from fits of $\chi$ which take
into account the analytic corrections. As final result we quote
\begin{equation}
\eta = 0.177(2),
\end{equation} 
where the error is such to include the estimates of $\eta$ obtained from the
analysis of $\ln Z$.  This result is consistent with the estimate $\eta =
0.180(5)$ reported in Ref.~\cite{HPPV-08}, but significantly more precise.

The results obtained here allow us to predict the behavior of the different
transition lines close to the MNP.  Standard scaling arguments predict that,
close to the MNP, the transition lines are given by
\begin{equation}
u_1 |u_2|^{-\phi} = X_{\pm},
\label{scaltc}
\end{equation}
where $X_{+}$ and $X_-$ are two constants that refer to the lines which
satisfy $T> T^*$ and $T<T^*$, respectively.  They can be determined by
considering the estimates of the critical points $p_c,T_c$ close to the MNP.
The crossover exponent $\phi$ is equal to the ratio $y_1/y_2$.  In the present
case we have
\begin{equation}
\phi = {y_1\over y_2} = 2.64(5).
\end{equation}
We can use Eq.~(\ref{scaltc}) to obtain an interpolation 
of our results up to $T=0$, which represents our best guess
of the transition line, given the estimates of the critical points we have.
For this purpose, we choose
\begin{eqnarray}
u_{2}(p,T) = \tanh (1/T) - 2 p + 1, 
\end{eqnarray}
so that $u_2=0$ along the N line, cf. Eq.~(\ref{tn}).
Thus, the critical line is given by the approximate expression
\begin{equation}
p_c - p^* + a_2 (T_c-T^*)^2 = X_- u_2(p_c,T_c)^\phi,
\label{interp-sotto}
\end{equation}
where we have kept the $O(\Delta T^2)$ in the analytic quadratic corrections
to the linear behavior of $u_1$, cf. Eq.~(\ref{u1def}).  Since $2<\phi<3$ this
quadratic term is dominant in the asymptotic expansion at the MNP, while the
nonanalytic term in the right-hand side of Eq.~(\ref{interp-sotto}) 
represents a next-to-leading
contribution. Since $p_c - p^* \sim (T_c-T^*)^2$ the other quadratic terms 
appearing in the expansion of $u_1$ are subleading.
The free parameters $a_2$ and
$X_-$ are fixed by requiring the line to go through the points
$(p_c=0.8925(1),T_c=0.5)$ and $(p_0=0.897,T=0)$.  We obtain $a_2=-0.0061$ and
$X_-=0.0386$.  The corresponding line is reported (dashed line) in
Fig.~\ref{phadiatc}.  The interpolation (\ref{interp-sotto}) gives
$p_c=0.89159$ at $\beta=1.55$, and the derivative $dp_c/d\beta = 0.00180$ at
$\beta=2$ which are in good agreement with the MC estimates $p_c=0.8915(2)$ at
$\beta=1.55$, and $dp_c/d\beta = 0.0020(3)$ at $\beta=2$ obtained in
Sec.~\ref{MCsec}.

We have also determined an interpolation of the available numerical data
\cite{MC-02,HPPV-08-2}
valid for $T>T^*$. A simple expression, which satisfies Eqs.~(\ref{scaltc})
and (\ref{TCder}), is 
\begin{equation}
p_c = p^* + (\beta^* - \beta)^{2.64}
    (1.41484 - 4.25764 \beta + 5.67965 \beta^2  - 2.77095 \beta^3),
\end{equation}
with $p^* = 0.89083$ and $\beta^* = 1.04962$. 
The corresponding line is reported (dashed line) in Fig.~\ref{phadiatc}.

\end{document}